\documentclass[pre,twocolumn,preprintnumbers,amsmath,amssymb]{revtex4}
\usepackage{amssymb}
\usepackage{latexsym}
\usepackage{epsfig}
\usepackage{color}

\begin{document}
\title{Implications of the Generalized Entropy Formalisms on the Newtonian Gravity and Dynamics}

\author{H. Moradpour$^1$\footnote{h.moradpour@riaam.ac.ir}, A. Sheykhi$^{2}$\footnote{asheykhi@shirazu.ac.ir}, C. Corda$^{1,3}$\footnote{cordac.galilei@gmail.com}, Ines G. Salako$^{4,5}$\footnote{inessalako@gmail.com}}
\address{$^1$ Research Institute
for Astronomy and Astrophysics of Maragha (RIAAM), P.O. Box
55134-441, Maragha, Iran\\
$^2$ Physics Department and Biruni Observatory, College of
Sciences, Shiraz University, Shiraz 71454,
Iran\\
$^3$ International Institute for Applicable Mathematics and Information Sciences, Adarshnagar, Hyderabad 500063 (India)\\
$^4$ D\'epartement de Physique, Universit\'e Nationale
d'Agriculture, 01 BP 55 Porto-Novo, Benin\\
$^5$ Institut de Math\'ematiques et de Sciences Physiques (IMSP),
Universit\'e  d'Abomey-Calavi Porto-Novo, 01 BP 613 Porto-Novo,
Benin}

\begin{abstract}
Employing the Verlinde's hypothesis, and considering two
well-known generalized entropy formalisms, two modifications to
the Newtonian gravity are derived. In addition, it has been shown
that the generalized entropy measures may also provide theoretical
basis for the Modified Newtonian Dynamics (MOND) theory and
generate its modified forms. Since these entropy measures are also
successful in describing the current accelerated universe, our
results indicate that the origin of dark sectors of cosmos may be
unified to meeting the generalized entropy measures instead of the
Boltzmann-Gibbs entropy by the gravitational systems due to the
long-range nature of gravity.
\end{abstract}
\maketitle
\section{Introduction\label{Intr}}

The correspondence between the first law of thermodynamics on the
boundary of spacetime and the field equations of gravity,
describing the system evolution in bulk
\cite{jacob,Pad1,jacob1,Pad0}, suggests a profound connection
between the laws of thermodynamics and gravity which further
supports the holography proposal \cite{Pad00,Pad11,Ver,cai,msrw}.
Based on the Verlinde's hypothesis \cite{Ver}, the tendency of
systems to increase their entropy leads to the emergence of
gravity between the holographic screens. It is important to note
that the entropy expression plays a crucial role in this theory.
In fact, diverse corrections to the entropy-area relation presents
various modifications to the gravitational theories and their
corresponding cosmology
\cite{Sheykhi1,Cai41,Li,Tian,Sheykhi2,Sheykhi21,Ling,other,mann,SMR,MS,nonex1,nonex2,zhang,SK,Gao,EN,EP,mas}.
It has also been shown \cite{zhang,SK,Gao,EN,EP} that if the Unruh
temperature \cite{unr} is attributed to the holographic screen,
then the quantum statistics of surface degrees of freedom may help
us in obtaining a theoretical basis for the Modified Newtonian
Dynamics (MOND) theory \cite{Milgrom,Milgrom1,Milgrom2}. It is
also worthy to note here that the quantum aspects of surface
degrees of freedom imply that $a_0$, appeared in MOND theory,
depends on the energy content of system
\cite{zhang,SK,Gao,EN,EP,Sheykhi2}.

One property of gravity is its long-range nature which motivates
physicists to use generalized entropy formalisms
\cite{1,fon,reyo1,abe,nn1,nn2,pla} in order to study various
gravitational and cosmological phenomena
\cite{nonex1,nonex2,non2,non19,non20,non22,non5,non6,non11,non14,eb1,
3,5,cite1,cite2,smm,me,ijtp,tskm,non3}. In the generalized entropy
formalisms, systems are described by the power-law distributions
of probabilities ($P_i^q$) instead of the ordinary linear
distribution ($P_i$) \cite{1,fon,reyo1,abe,nn1,nn2,pla}, and the
additional new free parameters, such as $q$, can be evaluated by
fitting with data \cite{1,fon,reyo1,abe,nn1,nn2,pla}. The Tsallis
generalized entropy \cite{1,fon}, an one-free parameter measure,
can be combined with the Verlinde's hypothesis \cite{Ver} to
obtain a MOND theory \cite{eb1}. Additionally, using the Verlinde
approach, it has also been shown that the power-law and
logarithmic corrections to the Bekenstein entropy lead to modified
versions for the MOND theory \cite{Sheykhi2}. These attempts
motivate us to study relations between those generalized entropies
which are successful in describing cosmological and gravitational
phenomena
\cite{nonex1,nonex2,non2,non19,non20,non22,non5,non6,non11,non14,eb1,
3,5,cite1,cite2,smm,me,ijtp,tskm,non3} and various MOND theories.

In the present Letter, by taking the various generalized entropy
formalisms as well as the entropic origin of gravity into account,
we are going to derive some MOND theories and the implications of
these generalized entropy measures on the Newtonian gravity. We
are focusing on those the generalized entropy measures successful
in describing the current accelerated universe
\cite{ijtp,non19,non20,smm,me,tskm}. The paper is organized as
follows. Some generalized entropy measures and the general
relation between the system entropy and the gravitational force in
the Verlinde approach have been shown in the next section. In
sec.~\textmd{III}, we address some MOND theories and corrected
Newtonian gravities based on the generalized entropies. The last
section is devoted to a summary. For the sake of simplicity, we
set $G=\hbar=c=k_B=1$, where $k_B$ is the Boltzmann constant,
throughout the article.
\section{Generalized entropy formalism, Verlinde hypothesis and the gravitational force}

For a system with $W$ discrete states, where each state has
probability $P_i$, Tsallis and R\'{e}nyi entropies are defined as

\begin{eqnarray}\label{reyn01}
S_T=\frac{1}{1-q}\sum_{i=1}^{W}\left(P_i^q-P_i\right),
\end{eqnarray}

\noindent and

\begin{eqnarray}\label{reyn02}
\mathcal{S}=\frac{1}{1-q}\ln\sum_{i=1}^{W} P_i^q,
\end{eqnarray}

\noindent respectively, where $q$ is a free parameter
\cite{reyo1}. Using the fact that $\sum P_i=1$, we can combine
these equations to arrive at

\begin{eqnarray}\label{reyn03}
(1-q)S_T+1=e^{(1-q)\mathcal{S}},
\end{eqnarray}

\noindent which finally leads to \cite{non2}

\begin{eqnarray}\label{reyn1}
\mathcal{S}=\frac{1}{\delta}\ln(1+\delta S_T),
\end{eqnarray}

\noindent where we have use the $\delta=1-q$ expression. There is
also a two-parametric generalized entropy which is called the
Sharma-Mittal entropy and is written as \cite{pla,smm}

\begin{eqnarray}\label{SME2}
S_{SM}=\frac{1}{\alpha}\big((1+\delta S_T)^{\frac{\alpha}{\delta}}-1\big),
\end{eqnarray}

\noindent where $\alpha\equiv1-r$, and $r$ is a new free
parameter. Some cosmic applications of $\mathcal{S}$ and $S_{SM}$
can be found in Refs. \cite{smm,me,non19,non20}. In general, the
free parameters $r$ and $q$ should be evaluated by comparing the
theory with the observations \cite{pla}, meaning that the free
parameters are not the same for all systems. This is in full
agrement with gravitational and cosmological studies
\cite{non2,non19,non20,non22,non5,non6,nonex2,non11,non14,eb1,
3,5,cite1,cite2,smm,me,ijtp,tskm,non3}.

In addition, we assume that system has a boundary with
area $A$, and consists $N$ degrees of freedom which satisfy the
energy equipartition law \cite{eb1,non19}

\begin{eqnarray}\label{CEPL}
E=M=\frac{NT}{2}.
\end{eqnarray}

\noindent Here, $T$ is the boundary temperature and $M$ denotes
the mass content of system. Moreover, in our unit, $A$ and
$N$ are in a mutual relation as \cite{Ver,Sheykhi21}

\begin{eqnarray}\label{N}
A=N,
\end{eqnarray}

\noindent claiming that the area change per unit change of
information is one (or equally $\Delta A=\Delta N=1$)
\cite{Ver,Sheykhi21}. It was argued that the Bekenstein's entropy
expression ($S_B={A}/{4}$), the entropy of a system with boundary
$A$ \cite{sered}, is a proper candidate for $S_T$
\cite{non2,non22,non19}, a result confirmed by using the Tsallis
formalism to evaluate the black hole entropy in loop quantum
gravity \cite{5}. Thus, replacing $S_B$ with $S_T$ in the above
generalized entropy measures, one can easily find the entropy-area
relation in generalized entropy formalism
\cite{smm,me,non19,non20}.

Based on the entropic force scenario, the absolute value of
gravitational force applied from a source $M$ to the test particle
$m$ located at the distance $\Delta x=\eta\lambda_m$ from the
holographic screen of radius $R$ covering $M$, is evaluated as
\cite{Ver,Sheykhi21}

\begin{eqnarray}\label{f}
F=T\frac{\Delta S}{\Delta x}.
\end{eqnarray}

\noindent Now, since $\frac{\Delta S}{\Delta x}=\frac{\Delta
S}{\Delta A}\frac{\Delta A}{\Delta
x}=(\frac{1}{\eta\lambda_m})\frac{dS}{dA}$ \cite{Ver,Sheykhi21},
using Eqs.~(\ref{CEPL}) and~(\ref{N}), this equation can be
written as

\begin{eqnarray}\label{f1}
F=(\frac{1}{2\pi\eta})\frac{Mm}{R^2}\frac{dS}{dA},
\end{eqnarray}

\noindent where $A=4\pi R^2$. We also used the Compton wavelength
expression $\lambda_m={1}/{m}$, as well as Eq.~(\ref{N}) to obtain
Eq.~(\ref{f1}).
\section{Possible MOND theories}

As we mentioned, the R\'{e}nyi and Sharma-Mittal entropies can
provide suitable description for the current accelerated universe
and thus dark energy \cite{non20,smm,me,non19}. Here, using the
introduced generalized entropy formalisms, Eq.~(\ref{N}), the
$S_B=S_T$ relation \cite{5}, and Eq.~(\ref{f1}), we are going to
get the various MOND theories allowed by employing the R\'{e}nyi
and Sharma-Mittal entropies to the system.
\subsection{R\'{e}nyi Entropy}

Now, inserting Eq.~(\ref{reyn1}) into equation (\ref{f1}), one can
easily obtain

\begin{eqnarray}\label{f2}
F=(\frac{1}{8\pi\eta})\frac{Mm}{R^2}\frac{1}{\delta A/4+1}.
\end{eqnarray}

\noindent Since we set $G=1$, we should have $F=\frac{Mm}{R^2}$ at the $\delta=0$ limit leading to $8\pi\eta=1$ in full
agreement with \cite{Ver,Sheykhi21}. Finally, defining
$\mathcal{A}_0\equiv\delta\pi M$ and
$a\equiv{M}/{R^2}$, one can rewrite Eq.~(\ref{f2}) as

\begin{eqnarray}\label{f3}
F=\left(\frac{1}{1+{\mathcal{A}_0}/{a}}\right)ma=f(a)ma,
\end{eqnarray}

\noindent which recovers the Newtonian gravity at the appropriate
limit of $\delta\rightarrow0$ (or equally
$\mathcal{A}_0\rightarrow0$). In addition, for $a\gg\mathcal{A}_0$
($a\ll\mathcal{A}_0$), we have $F\simeq ma={Mm}/{R^2}$ ($F\simeq
m\frac{a^2}{\mathcal{A}_0}={Mm}/{\pi\delta R^4}$) meaning that
the Modified Newtonian force obtained in Eq.~(\ref{f3}) has
similarities with a MOND theory with simple interpolating function
$\mu\left(\frac{\mathcal{A}_0}{a}\right)={1}/{(1+{\mathcal{A}_0}/{a})}$.

From the phenomenological point of view, following \cite{92,93} and working with the accelerations, one can introduce vectors and
divides Eq.~(\ref{f3}) as

\begin{equation}
\overrightarrow{a}=\overrightarrow{a}_{n}+\overrightarrow{a}{}_{nn},\label{eq: vect}
\end{equation}

\noindent where the total acceleration $\overrightarrow{a}$ is
given by the ordinary Newtonian acceleration
$\overrightarrow{a}_{n}$ plus the acceleration
$\overrightarrow{a}_{nn}$ which is due to the non-Newtonian force.
Now, taking the square of Eq. (12) one gets [59]:

\begin{equation}
\vec{a}_{nn}\cdot\vec{a}_{n}=\frac{1}{2}\left(a^{2}-a_{n}^{2}-a_{nn}^{2}\right).\label{eq: 13}
\end{equation}

\noindent In this equation the dot represents the
three-dimensional scalar product. Equation (13) is a general
relation which expresses the unknown vector $\vec{a}_{n}$ in terms
of the total acceleration $\vec{a}$, of the acceleration due to
the non-Newtonian force $\vec{a}_{nn}$ and of the magnitudes
$a^{2},$ $a_{n}^{2}$ and $a_{nn}^{2}$. From Eq. (13), one obtains
the acceleration $\vec{a}_{n}$ as being [59]

\begin{equation}
\overrightarrow{a}_{n}=\frac{1}{2}(a^{2}-a_{n}^{2}-a_{nn}^{2})\frac{\overrightarrow{a}}{a_{nn}\cdot a}+\vec{b}\,\mathbb{\mathrm{x}}\,\vec{a}_{nn},\label{eq:  14}
\end{equation}

\noindent where $\vec{b}$ is is an arbitrary vector perpendicular
to the acceleration $\vec{a}_{nn}.$ For the sake of simplicity,
one assumes $\vec{b}=0$ [59]. As one wants the mathematical
consistency of Eq. (14), one needs $a_{nn}\cdot a\neq0$. In other
words, the accelerations $a_{nn}$ and $a$ cannot be orthogonal to
each other. We will assume that both of them are parallel. Thus,
Eq. (\ref{eq:  14}) becomes

\begin{equation}
\overrightarrow{a}_{n}=\frac{1}{2}(a^{2}-a_{n}^{2}-a_{nn}^{2})\frac{\overrightarrow{a}}{a_{nn}a}.\label{eq: Newt}
\end{equation}

\noindent Assuming that $\overrightarrow{a}{}_{ng}$ dominates, which means
$a_{n}\ll a,$ one gets

\begin{equation}
a_{n}\simeq\frac{a\overrightarrow{a}}{2a_{nn}}(1-\frac{a_{nn}^{2}}{a^{2}}).\label{eq: a1}
\end{equation}

\noindent If one defines \cite{92,93}

\begin{equation}
a_{e}^{-1}\equiv\frac{1}{2a_{nn}}(1-\frac{a_{n}^{2}}{a^{2}}),\label{eq: aE}
\end{equation}

\noindent then Eq. (\ref{eq: a1}) becomes

\begin{equation}
\overrightarrow{a}_{n}\simeq\frac{a}{a_{e}}\overrightarrow{a},\label{eq:
a1 due}
\end{equation}

\noindent which can be combined with Eq. (\ref{eq: a1 due}) to obtain

\begin{equation}
a\simeq(a_{e}\textrm{}a_{n})^{\frac{1}{2}}.\label{eq: a tot 2}
\end{equation}

We now recall that the standard Newtonian acceleration is

\begin{equation}
\overrightarrow{a}_{n}=\frac{M}{r^{2}}\widehat{u}_{r},\label{eq: StNe}
\end{equation}

\noindent helping us in writing the total acceleration as

\begin{equation}
\overrightarrow{a}=\frac{(a_{e}M)^{\frac{1}{2}}}{r}\widehat{u}_{r}=\frac{v_{r}^{2}}{r}\widehat{u}_{r},\label{eq: a tot rad}
\end{equation}

\noindent where

\begin{equation}
v_{r}=(a_{e}M)^{\frac{1}{4}},\label{eq: vr}
\end{equation}

\noindent is the rotation velocity of a test mass under the influence of the
non-Newtonian force. By applying our analysis to the galaxies rotation curves, one can
identify in a natural way $a_{e}$ with $a_{0}\simeq10^{-10}m/s^2$,
which is analogous to the Milgrom's MOND acceleration \cite{Milgrom,Milgrom1,Milgrom2}.


It has been argued that since the holographic screen, assumed in previous section, can be
considered as the system boundary \cite{Ver,zhang,SK}, one can attribute
Unruh temperature \cite{unr}

\begin{eqnarray}\label{un}
T=\frac{a}{2\pi},
\end{eqnarray}

\noindent to this boundary \cite{Sheykhi1,Cai41,Ver,zhang,SK,eb1}.
Combining this equation with Eqs.~(\ref{CEPL}) and~(\ref{N}), we
get

\begin{eqnarray}\label{Aa}
A=\frac{4\pi E}{a},
\end{eqnarray}

\noindent as a relation between the surface area of holographic
screen and the acceleration of test particle. This relation
indicates that $A$ is decreased by increasing $a$. It is a true
result because attracting the test particle by the source, their
mutual interval and thus the holographic screen are shrinking.

Now, bearing the $\frac{\Delta S}{\Delta
x}=(\frac{1}{\eta\lambda_m})\frac{dS}{dA}$ and
$\lambda_m=\frac{1}{m}$ relations in mind, by inserting
Eqs.~(\ref{un}) and~(\ref{Aa}) into Eq.~(\ref{f}) and using
Eq.~(\ref{reyn1}), one can find

\begin{eqnarray}\label{mond1}
F=(\frac{1}{8\pi\eta})\frac{ma}{1+\frac{a_0}{a}},
\end{eqnarray}

\noindent in which

\begin{eqnarray}\label{mond2}
a_0\equiv\delta\pi M.
\end{eqnarray}

\noindent It is obvious that the Newtonian dynamics is obtainable
at the appropriate limit $\delta=0$ ($a_0=0$), whenever
$8\pi\eta=1$, a result in full accordance with what is obtained in
Eq.~(\ref{f2}). Therefore, in order to obtain a MOND theory with
simple interpolating function
$\mu\left(\frac{a_0}{a}\right)={1}/{(1+{a_0}/{a})}$, we should
have $\eta=1/8\pi$ which leads to $a_0=\delta\pi M$. Thus
$a_0=\mathcal{A}_0$, and the amount of $a_0$ depends on both of
$\delta$ and the mass content of source.

It is worth to mention here that the linear dependency of $a_0$ to
the mass (energy) content of system, obtained in
Eq.~(\ref{mond2}), is in full agreement with the results of
applying quantum statistics to the surface degrees of freedom
\cite{EN,EP,Sheykhi2}. This result also says that $\delta$ should
be evaluated by comparing the observation with Eq.~(\ref{mond2}),
and its value is not necessarily the same for all the galaxies.
For example, if $a_0\simeq10^{-10}m/s^2$
\cite{Milgrom,Milgrom1,Milgrom2,SK}, then we can use
Eq.~(\ref{mond2}) in order to find $\delta$ (or equally $q$) for a
system with mass $M$. It is in agreement with the spirit of
generalized entropy formalism in which the values of free
parameters should be evaluated by fitting the theory with the
observations \cite{1,fon,reyo1,abe,nn1,nn2,pla}.

\subsection{Sharma-Mitall Entropy}

Employing the $S_B=S_T=\frac{A}{4}$ relation \cite{5} and
Eq.~(\ref{N}), one can rewrite Eq.~(\ref{SME2}) as

\begin{eqnarray}\label{SME3}
S_{SM}=\frac{1}{\alpha}\big((1+\frac{\delta
A}{4})^{\frac{\alpha}{\delta}}-1\big).
\end{eqnarray}

\noindent It is the generalization of both the Tsallis and
R\'{e}nyi entropy measures \cite{pla} which reduces to the
Bekenstein entropy when $\alpha=\delta$ \cite{smm}. Now, following
the approaches led to Eqs.~(\ref{f2}) and~(\ref{mond1}), we reach

\begin{eqnarray}\label{smf}
F=\big(\frac{Mm}{R^2}\big)[\frac{1}{\frac{\delta A}{4}+1}]^{1-\frac{\alpha}{\delta}},
\end{eqnarray}

\noindent and

\begin{eqnarray}\label{smmond}
F=ma[\frac{1}{\frac{a_0}{a}+1}]^{1-\frac{\alpha}{\delta}},
\end{eqnarray}

\noindent respectively, where $a_0$ meets Eq.~(\ref{mond2}). In
fact, since the Newtonian gravity and dynamics are obtainable
whenever the Bekenstein entropy is considered \cite{Sheykhi2}, our
results should reduce to those of the Newton at the
$\alpha=\delta$ limit. This expectation is obeyed whenever
$\alpha=\delta$ leading to $S_{SM}=S_B$ \cite{smm} and
$8\pi\eta=1$ used to obtain the above equations.

Let us focus on the last equation indeed a MOND-like theory with
the interpolating function
$\xi(\frac{a_0}{a})=\frac{1}{(\frac{a_0}{a}+1)^{1-\frac{\alpha}{\delta}}}$.
In this manner, we have

\begin{eqnarray}\label{xi1}
\xi(\frac{a_0}{a})=\bigg\{^{1,\ a\gg a_0 \ (independent\ of\ the\
values\ of\ \delta\ and\
\alpha),}_{\bigg\{^{(\frac{a}{a_0})^{1-\frac{\alpha}{\delta}}\
for\
\frac{\alpha}{\delta}<1}_{{(\frac{a_0}{a})^{\frac{\alpha}{\delta}-1}\
for\ \frac{\alpha}{\delta}>1}}\ a_0\gg a,}
\end{eqnarray}

\noindent which finally leads to

\begin{eqnarray}\label{xi2}
F=\bigg\{^{ma,\ a\gg a_0 \ (independent\ of\ the\ values\ of\
\delta\ and\
\alpha),}_{\bigg\{^{ma(\frac{a}{a_0})^{1-\frac{\alpha}{\delta}}\
for\
\frac{\alpha}{\delta}<1}_{{ma(\frac{a_0}{a})^{\frac{\alpha}{\delta}-1}\
for\ \frac{\alpha}{\delta}>1}}\ a_0\gg a,}
\end{eqnarray}

\noindent for the force felt by a particle with mass $m$ and
acceleration  $a$. Therefore, the Sharma-Mitall entropy modifies
the MOND theory as Eq.~(\ref{xi2}). Similar modifications to the
MOND theory have also been obtained by using the logarithmic and
power-law corrections of entropy \cite{Sheykhi2}.
\section{Summary}

Bearing the entropic force scenario in mind, we studied some
consequences of applying two generalized entropy formalisms,
successful in describing the current accelerated universe
\cite{non19,non20,smm,me,tskm}, to the gravitational systems. It
was shown that the R\'{e}nyi entropy modifies the Newtonian
gravity. Moreover, we also found out that this entropy measure can
provide a theoretical basis for the MOND theory with the simple
interpolating function
$\mu\left(\frac{a_0}{a}\right)=\frac{1}{1+{a_0}/{a}}$. The
Sharma-Mitall entropy has also been studied showing that this
entropy can modify both the Newtonian gravity and the MOND theory.
Our results express that, the same as the dark energy origin
\cite{non19,non20,smm,me,tskm}, the nature of MOND theory may be
attributed to the long-range aspect of gravity which may force the
gravitational systems to obey the generalized entropy formalisms
instead of the ordinary Boltzmann-Gibbs entropy.

\acknowledgments{The authors thank the anonymous referee for valuable comments. We are so grateful to Prof. Nobuyoshi Komatsu for
valuable tips. The work of H. Moradpour has been supported
financially by Research Institute for Astronomy \& Astrophysics of
Maragha (RIAAM).}



\begin{thebibliography}{99}
\bibitem{jacob} T. Jacobson, Phys. Rev. Lett. \textbf{75}, 1260 (1995).
\bibitem{Pad1} T. Padmanabhan, Phys. Rep. 406, 49 (2005).
\bibitem{jacob1} C. Eling, R. Guedens, T. Jacobson, Phys. Rev. Lett. {\bf96}, 121301 (2006).
\bibitem{Pad0} T. Padmanabhan, Class. Quantum. Grav. {\bf19}, 5387 (2002).
\bibitem{Pad11} T. Padmanabhan, Rep. Prog. Phys. 73, 046901 (2010).
\bibitem{Pad00} T. Padmanabhan, arXiv:0910.0839 (2009).
\bibitem{Ver} E. Verlinde, JHEP {\bf1104}, 029 (2011).
\bibitem{cai} R. G. Cai, L. M. Cao, N. Ohta, Phys. Rev. D {\bf81}, 061501(R) (2010).
\bibitem{msrw} H. Moradpour, A. Sheykhi, N. Riazi, B. Wang, AHEP, 718583 (2014).
\bibitem{Sheykhi1} A. Sheykhi, Phys. Rev. D \textbf{81}, 104011 (2010).
\bibitem{Cai41} R. G. Cai, L. M. Cao, N. Ohta, Phys. Rev. D {\bf81}, 084012 (2010).
\bibitem{Li} M. Li, Y. Wang, Phys. Lett. B {\bf687}, 243 (2010).
\bibitem{Tian} Y. Tian, X. Wu, Phys. Rev. D \textbf{81}, 104013 (2010).
\bibitem{Sheykhi2} S. H. Hendi, A. Sheykhi, Phys.  Rev.  D {\bf83}, 084012 (2011).
\bibitem{Sheykhi21} A. Sheykhi, S. H. Hendi, Phys.  Rev.  D {\bf84}, 044023 (2011).
\bibitem{Ling} Y. Ling, J. P. Wu, J. Cosmol. Astropart. Phys. {\bf1008}, 017 (2010).
\bibitem{other} R. C. Nunes, et al. Int. J. Geo. Meth. Mod. Phys. 15, 1850004 (2018).
\bibitem{mann} R. B. Mann, J. R. Mureika, Phys. Lett. B {\bf703}, 167 (2011).
\bibitem{SMR} A. Sheykhi, H. Moradpour, N. Riazi, Gen. Rel. Grav. {\bf45}, 1033 (2013).
\bibitem{MS} H. Moradpour, A. Sheykhi, Int. J. Theo. Phys. {\bf55}, 1145 (2016).
\bibitem{nonex1} E. M. Barboza Jr., R. C. Nunes, E. M. C. Abreu, J. Ananias Neto, Phys. Lett. B 436, 301 (2015).
\bibitem{nonex2} R. C. Nunes, et al. JCAP, 08, 051 (2016).
\bibitem{zhang} H. Zhang, X. Z. Li, Phys. Lett. B 715, 15 (2012).
\bibitem{SK} A. Sheykhi, K. Rezazadeh Sarab, J. Cosmol. Astropart. Phys. {\bf10}, 012 (2012).
\bibitem{Gao} C. Gao, Phys. Rev. D {\bf81}, 087306 (2010).
\bibitem{EN} E. Pazy, N. Argaman, Phy. Rev. D {\bf85}, 104021 (2012).
\bibitem{EP} E. Pazy, Phy. Rev. D {\bf87}, 084063 (2013).
\bibitem{mas} H. Moradpour, A. Amiri, A. Sheykhi, Iran. J. Sci. Technol. Trans. Sci. (2018) DOI: 10.1007/s40995-018-0569-x [arXiv:1710.07224]
\bibitem{unr} W. G. Unruh, Phys. Rev. D {\bf14}, 870 (1976).
\bibitem{Milgrom} M. Milgrom, Astrop. J 270, 365 (1983).
\bibitem{Milgrom1} M. Milgrom, Astrop. J 270, 371 (1983).
\bibitem{Milgrom2} M. Milgrom, Astrop. J 270, 384 (1983).
\bibitem{1} M. Gell-Mann, C. Tsallis, \textit{Nonextensive Entropy-Interdisciplinary Applications}, (Oxford University Press, New York, 2004).
\bibitem{fon} S. Abe, \textit{Foundations of Nonextensive Statistical Mechanics. In: Sengupta A. (eds) Chaos, Nonlinearity, Complexity. Studies in Fuzziness and Soft Computing} 206. (Springer, Berlin, Heidelberg, 2006).
\bibitem{reyo1} A. R\'{e}nyi, \textit{Probability Theory} (North-Holland, Amsterdam, 1970).
\bibitem{abe} S. Abe, Phys. Rev. E 63, 061105 (2001).
\bibitem{nn1} H. Touchette, Physica A 305, 84 (2002).
\bibitem{nn2} T. S. Bir\'{o}, P. V\'{a}n, Phys. Rev. E 83, 061147 (2011).
\bibitem{pla} M. Masi, Phys. Lett. A 338, 217 (2005).
\bibitem{non3} C. Tsallis, L. J. L. Cirto, Eur. Phys. J. C 73, 2487 (2013).
\bibitem{non2} T. S. Bir\'{o}, V. G. Czinner, Phys. Lett. B 726, 861 (2013).
\bibitem{non22} V. G. Czinner, H. Iguchi, Phys. Lett. B 752, 306 (2016).
\bibitem{non11} W. Guo, M. Li, Nucl. Phys. B 882, 128 (2014).
\bibitem{non5} E. M. C. Abreu, J. Ananias Neto, Phys. Lett. B 727, 524 (2013).
\bibitem{non6} E. M. Barboza Jr., R. C. Nunes, E. M. C. Abreu, J. A. Neto, Physica A: Statistical Mechanics and its Applications, 436, 301 (2015).
\bibitem{non14} N. Komatsu, S. Kimura. Phys. Rev. D 93, 043530 (2016).
\bibitem{eb1} E. M. C. Abreu, J. A. Neto, A. C. R. Mendes, D. O. Souza, EPL 120, 20003 (2017).
\bibitem{3} A. Taruya, M. Sakagami, Phys. Rev. Lett. 90, 181101 (2003).
\bibitem{5} A. Majhi, Phys. Lett. B 775, 32 (2017).
\bibitem{cite1} V. G. Czinner, F. C. Mena, Phys. Lett. B 758, 9 (2016).
\bibitem{cite2} V. G. Czinner, H. Iguchi, Eur. Phys. J. C 77, 892 (2017).
\bibitem{ijtp} H. Moradpour, Int. J. Theor. Phys. 55, 4176 (2016).
\bibitem{non19} N. Komatsu, Eur. Phys. J. C 77, 229 (2017).
\bibitem{non20} H. Moradpour, A. Bonilla, E. M. C. Abreu, J. A. Neto, Phys. Rev. D 96, 123504 (2017).
\bibitem{smm} A. Sayahian Jahromi et al., Phys. Lett. B 780, 21 (2018).
\bibitem{me} H. Moradpour et al., arXiv:1803.02195v1.
\bibitem{tskm} M. Tavayef, A. Sheykhi, K. Bamba, H. Moradpour, Phys. Lett. B 781, 195 (2018).
\bibitem{sered} M. Srednicki, Phys. Rev. Lett. 71, 666 (1993).
\bibitem{92} O. Bertolami, C. G. Bohmer, T. Harko  and F. S. M. Lobo, Phys. Rev. D 75 104016 (2007)
\bibitem{93} C. Corda, Mod. Phys. Lett. A 23, 109 (2008).
\end{thebibliography}
\end{document}